\documentclass[twocolumn,amsmath,amssymb,aps,prc,nofootinbib]{revtex4-2}
\usepackage{graphicx,here}
\usepackage{dcolumn,bm,hyperref,footnote}
\usepackage{float}
\usepackage[dvipsnames,svgnames,x11names]{xcolor}
\usepackage[authormarkup=none,defaultcolor=BrickRed]{changes}
\usepackage{multirow}
\usepackage{lipsum}
\usepackage{tikz}
\usepackage{booktabs}

 \usepackage{bbold}
 \usepackage{subcaption}
\begin{document}

\title{Weighted Levenberg-Marquardt methods for fitting multichannel nuclear cross section data}

\author{M. Imbri{\v s}ak}
\email{marko.imbrisak@gmail.com}
\affiliation{Los Alamos National Laboratory, Los Alamos, NM, 87545, USA}
\author{A. E. Lovell}
\affiliation{Los Alamos National Laboratory, Los Alamos, NM, 87545, USA}
\author{M. R. Mumpower}
\affiliation{Los Alamos National Laboratory, Los Alamos, NM, 87545, USA}

\date{\today}

\begin{abstract}
We present an extension of the Levenberg–Marquardt algorithm for fitting multichannel nuclear cross section data. Our approach offers a practical and robust alternative to conventional trust-region methods for analyzing experimental data. The CoH$_3$ code, based on the Hauser–Feshbach statistical model, involves a large number of interdependent parameters, making optimization challenging due to the presence of ``sloppy” directions in parameter space.
To address the uneven distribution of experimental data across reaction channels, we construct a \emph{weighted Fisher Information Metric} by integrating prior distributions over dataset weights. This framework enables a more balanced treatment of heterogeneous data, improving both parameter estimation and convergence robustness.
We show that the resulting \emph{weighted Levenberg–Marquardt} method yields more physically consistent fits for both raw and smoothed datasets, using experimental data for ${}^{148}$Sm as a representative example. Additionally, we introduce a geometric scaling strategy to accelerate convergence--- a method based on the local geometry of the manifold.

\end{abstract}

\maketitle

\section{Introduction\label{sec:intro}}
{\color{black}
Fitting complex physical models to experimental data often leads to challenges due to parameter degeneracies and ill-conditioning in the optimization landscape. The Levenberg-Marquardt (LM) algorithm \citep{marquardt1963algorithm}, a widely used nonlinear least-squares minimizer, is known to struggle near parameter boundaries and can suffer from poor convergence behavior in models with many parameters \citep{Transtrum2010}. This difficulty arises because many physical models (e.g. in nuclear physics) are structurally ``sloppy" --- the eigenvalues of the Fisher Information Metric (FIM) differ logarithmically by many orders of magnitude \citep{Transtrum2012, 2024arXiv240300753P}. Since the inverse of the FIM provides an estimator for the parameter covariance matrix via the Cramér–Rao bound \citep{Cramer99}, such a wide eigenvalue spectrum complicates robust parameter estimation and uncertainty quantification. Parameters that are tightly constrained by the data are called ``stiff" parameters and they corresponding to large FIM eigenvalues. In contrast, parameters that are associated with small FIM eigenvalues are called sloppy. They are weakly constrained and contribute the most to the model’s uncertainty. 

We address these challenges by generalizing a weighting technique originally developed for cosmological data analysis \citep{Hobson2002}.  This weighting approach has been used to first construct a weighted Fisher Information Metric (wFIM) that was later applied in key steps of the classical LM technique. 
As it will be shown later in paper, this novel algorithm, that we have named weighted Levenberg-Marquardt (wLM) works by incorporating dataset-specific weights in a novel way that enables consistent and simultaneous fitting of multiple observables---where each observable is associated with a subset of data that may vary in both quantity (number of measurements) and quality (reported uncertainties). 

A trust region is a region of the parameter space where the model's linear or quadratic approximation of the objective function is presumed reliable \citep[e.g.,][]{berghen2004condor}. Instead of making unconstrained updates to parameters, trust-region methods restrict optimization steps to lie within a localized region of the parameter space, e.g., using reparametrizations \cite{James1975,James2004}. This region is dynamically updated based on how well the model's predicted improvement matches the actual reduction of the cost function. The wLM formulation enables us to navigate the complex topology of the parameter space, mitigate the impact of data imbalance---both in quantity and quality--- and impose trust-region-like constraints even in the absence of clearly defined parameter bounds. The wLM method facilitates the coherent treatment of heterogeneous datasets---such as cross section measurements from different reaction channels and experimental groups by introducing a flexible weighting prior. The wLM algorithm effectively incorporates a trust-region-like constraint by reweighting the influence of each dataset on parameter updates, thereby allowing for stable convergence without rigid parameter bounds.

In practical evaluations of nuclear data, the EXFOR database \citep{ENDF2014} is commonly used to aggregate experimental cross section data across many nuclei and reaction channels. However, experimental uncertainties---especially correlations between different datasets---are frequently not reported, potentially introducing bias in the fits. 
The Full Bayesian Evaluation Technique (FBET) \citep{leeb_consistent_2008,leeb_geneus_2011} addresses this issue by using an iterative generalized least-squares approach that provides a total estimate of a covariance matrix for a given energy grid, since it adequately accounts for both experimental and model-induced errors \citep{neudecker_advanced_2014,neudecker_impact_2013}. 
FBET smoothing
becomes necessary when working with larger datasets, as
it reduces the number of required model evaluations.

As a complementary effort, templates of expected experimental uncertainties have more recently been developed for key reaction observables---such as total, capture, and (n,xn) cross sections. These templates identify relevant sources of uncertainty for each measurement type and suggest typical ranges and correlation structures based on surveys of experimental data and expert input \citep{Neudecker2018,Neudecker2023}. These templates are intended to assist both EXFOR compilers and evaluators in achieving more realistic and consistent uncertainty quantification across datasets.

To explain experimental data, the Hauser-Feshbach (HF) formalism remains the standard approach for calculating average cross sections in the regime where compound nucleus resonances overlap. Modern HF codes such as \textsc{EMPIRE} \citep{INDC060}, \textsc{TALYS} \citep{Koning2012}, \textsc{CCONE} \citep{Iwamoto2007}, and \textsc{CoH}$_3$ \citep{Kawano2010,Kawano2019} incorporate additional physical mechanisms, including pre-equilibrium effects, $\gamma$-ray production, and multi-particle emission. Among these, \textsc{CoH}$_3$ is distinguished by its detailed treatment of nuclear deformation via coupled-channels optical model calculations, spanning a broad energy range from the keV scale to tens of MeV. It relies on a range of model parameters, particularly those modifying the standard Woods-Saxon potential, to encode nuclear structure effects.
While parameters such as optical model depths, radii, and diffusenesses are generally expected to remain close to physically motivated baseline values \cite{Koning2003,Becchetti1969,Varner1991}, their impact on observables can vary significantly due to channel coupling and interference effects.

The wLM algorithm we introduce provides a flexible approach for constrained fitting of optical model parameters in reaction codes such as \textsc{CoH}$_3$, capable of accommodating both raw experimental data and FBET-smoothed datasets. By incorporating weighting priors into the optimization process, wLM offers a principled framework for reconciling tensions between datasets and extracting robust model parameters.

In Section~\ref{sec:minimization}, we derive and implement the wLM algorithm. Section~\ref{sec:methods} details the FBET procedure and the structure of the \textsc{CoH}$_3$ model. Section~\ref{sec:results} presents numerical experiments demonstrating the effectiveness of our approach.
 
 }

\section{Minimization techniques}\label{sec:minimization}
In the next sections, we describe modified versions of the LM algorithm. Because different measurement techniques are used to collect data for different reaction channels and the distribution of data points in each channel may not cover the same energy range, the standard LM algorithm often provides the strongest constraint on the best-populated channel and disregards the rest.

Furthermore, there is also an issue of sloppy models, for which the eigenvalues of the covariance matrix span several orders of magnitude, indicating unconstrained parameters; this can lead to the LM-derived minimum being outside of the physically reasonable region of the parameter space. Classical trust-region methods, such as reparametrization \cite{James2004}, can result in a minimization procedure that converges at the trust-region's boundary, which often does not provide a good fit to the data. 
In this section, we derive the weighted Fisher information metric and use it to extend the classical LM algorithm. We use the Einstein summation convention throughout.
\subsection{Weighting the Fisher information metric\label{sec:model}}

We seek to analyze a dataset composed of multiple measurement groups, each representing either a distinct nuclear reaction channel or originating from a different experimental study. The measurements are thus grouped into $N_g$ distinct groups, $k\in\{1,\cdots,N_g\}$, each containing $n_k$ elements. We label a particular measurement as $y^{(k)}{}^i$, where $i$ indicates the i-th measurement in group $k$, $i\in \{1,\cdots, n_k\}$. Each measurement has its respective measurement error $\sigma^{(k)}{}^i$ and is described by the model $f^{(k)}{}^i(\theta)$, parametrized by $N_p$ parameters $\theta^\mu=\{\theta^1,\cdots,\theta^{N_p}\}$.
Using standard statistical assumptions, we define the residuals as
\begin{equation}
    r^{(k)}{}^{i}(\theta) = \frac{y^{(k)}{}^i - f^{(k)}{}^i(\theta)}{\sigma^{(k)}{}^i},
\end{equation}
which we assume to be independently and normally distributed, i.e., $r^{(k)}{}^{i}(\theta) \sim \mathcal{N}(0,1)$.

When measurements within each group exhibit correlations described by a covariance matrix $\Sigma^{(k)}{}^{ij}$, the residuals generalize to
\begin{equation}
    r^{(k)}{}^{i}(\theta) = \sum_{j=1}^{n_k} \left[(\Sigma^{(k)})^{-1/2}\right]^i_j \bigl(y^{(k)}{}^j - f^{(k)}{}^j(\theta)\bigr),
\end{equation}
where $(\Sigma^{(k)})^{-1/2}$ denotes a matrix square root of the inverse covariance matrix.

The corresponding \emph{group-specific} chi-squared statistic is then
\begin{equation}
    \chi^2_k(\theta) = \sum_{i=1}^{n_k} \bigl(r^{(k)}{}^{i}(\theta)\bigr)^2.
\end{equation}

\begin{figure*}
     \includegraphics[width=1.05\textwidth]{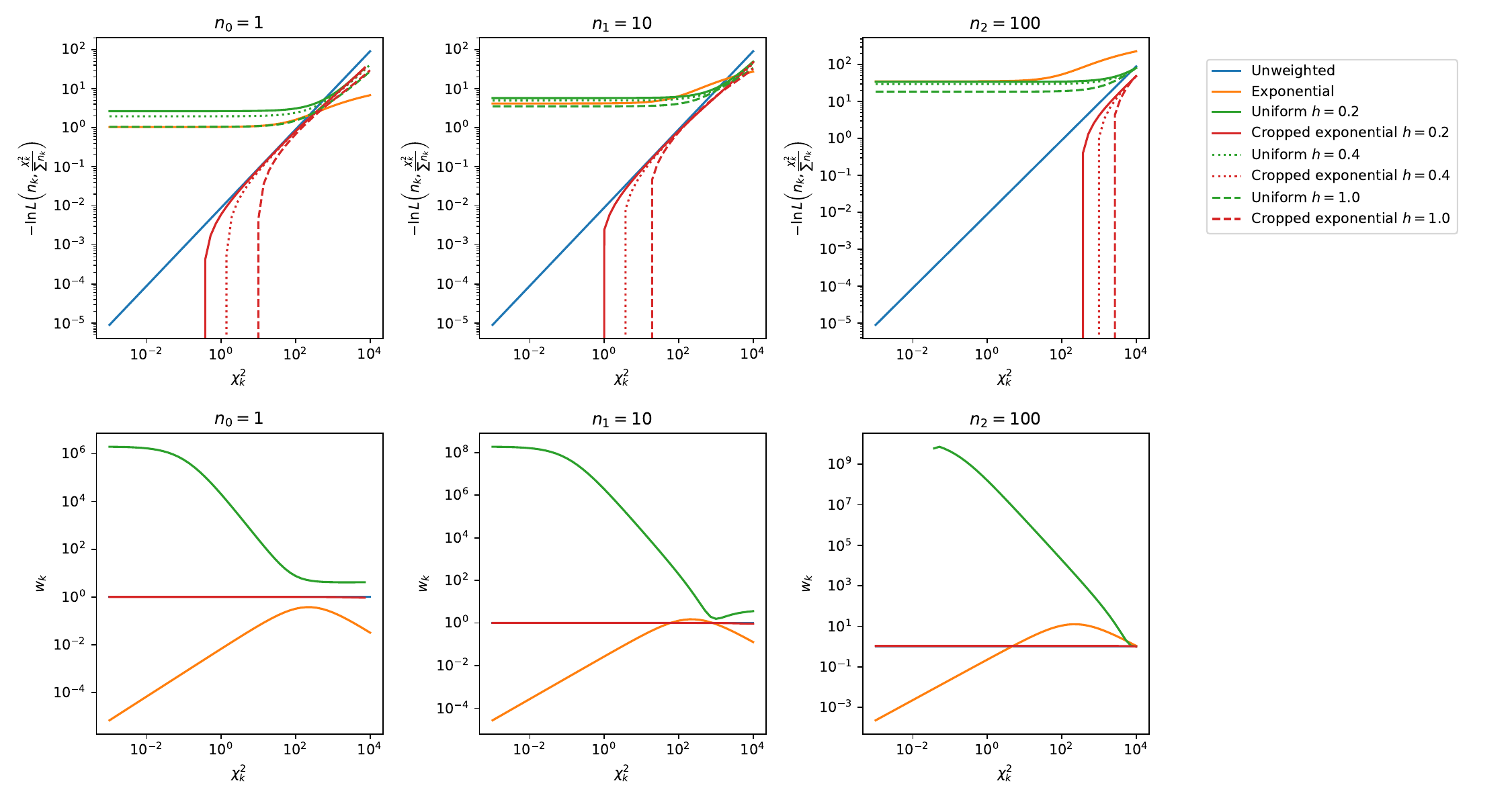}
     \caption{
Negative log-likelihoods (top row) and corresponding wFIM weights (bottom row) for the weighting schemes introduced in Section~\ref{sec:model}. The weights determine the relative contribution of each group to the total cost function during optimization. Results are shown for the unweighted (blue), exponential (orange), uniform (green), and cropped exponential priors (red). For the uniform and cropped exponential cases (solid, dotted, and dashed), three representative values of the width parameter $h$ are used to illustrate the effect of prior support. All curves are plotted as functions of $\chi^2_k$ for representative group sizes $n_k$.}
\label{fig:weights}
 \end{figure*}
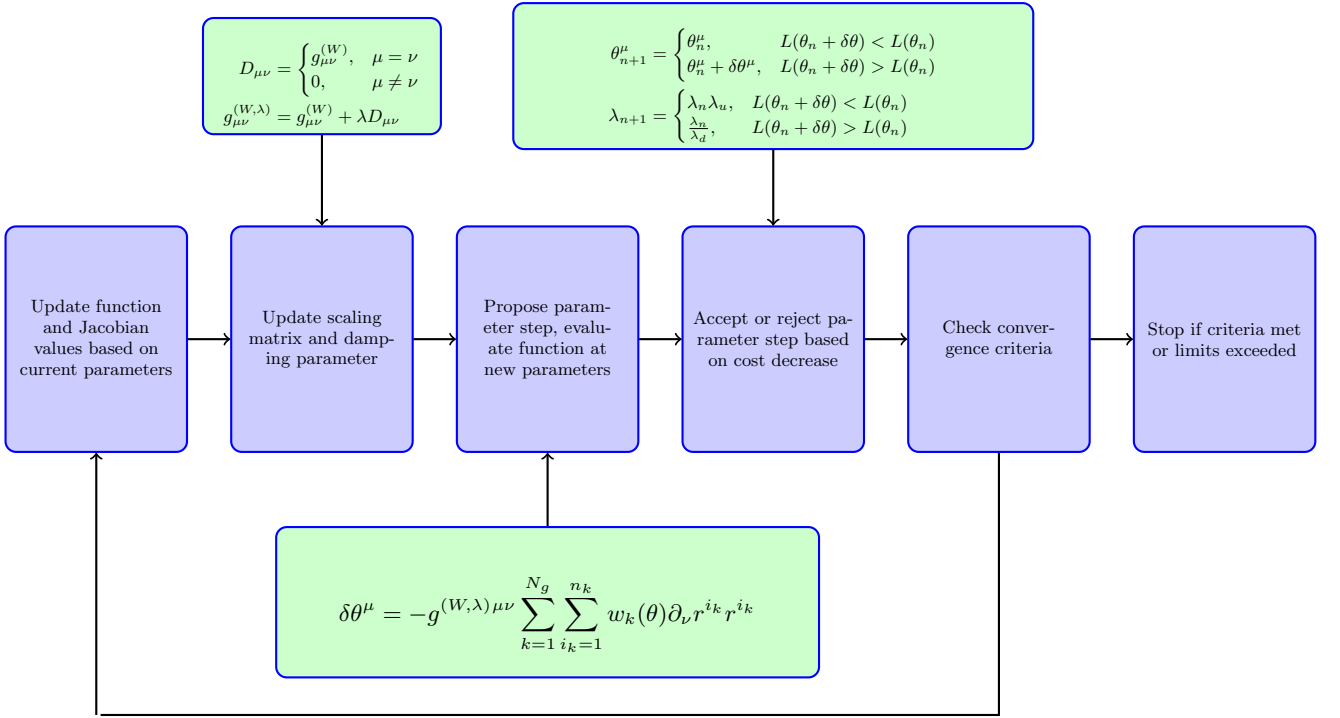
\begin{figure*}
\begin{tikzpicture}[scale=.5, transform shape,block/.style={
      rectangle,
      draw=blue,
      thick,
      fill=blue!20,
      align=center,
      rounded corners,
      minimum height=4cm,scale=1.5
    },block2/.style={
      rectangle,
      draw=blue,
      thick,
      fill=green!20,
      align=center,
      rounded corners,
      minimum height=2cm
    },block3/.style={
      rectangle,
      draw=blue,
      thick,
      fill=blue!20,
      align=center,
      rounded corners,
      minimum height=2cm
    }]
    \node[block,text width=3cm,align=center] (a) at (0,0)   {Update function 
    and Jacobian values based on current parameters}; 
    \node[block,text width=3cm,align=center] (b) at (6,0)   {Update scaling matrix and damping parameter}; 
    \node[block,text width=3cm,align=center] (c) at (12,0)   {Propose parameter step, evaluate function at new parameters}; 
    \node[block,text width=3cm,align=center] (d) at (18,0)   {Accept or reject parameter step based on cost decrease}; 
    \node[block,text width=3cm,align=center] (e) at (24,0)   {Check convergence criteria}; 
    \node[block,text width=3cm,align=center] (f) at (30,0)   {Stop if criteria met or limits exceeded}; 
    
    \node[block2,text width=4cm,align=center,scale=1.5] (g) at (6,7)   {%
    \begin{align*}
D_{\mu\nu}&=\begin{cases}g^{(W)}_{\mu\nu},& \mu=\nu\\
        0, & \mu\neq \nu\end{cases}\\
g_{\mu\nu}^{(W,\lambda)}&=g^{(W)}_{\mu\nu}+\lambda D_{\mu\nu}
    \end{align*}
    };
    \node[block2,text width=7cm,align=center,scale=2] (h) at (12,-7)   {%
    \begin{align*}
        \delta\theta^\mu=-g^{(W,\lambda)}{}^{\mu\nu}\sum\limits_{k=1}^{N_g}\sum\limits_{i_k=1}^{n_k} w_k(\theta)\partial_\nu r^{i_k} r^{i_k}
    \end{align*}
    };
    \node[block2,text width=9cm,align=center,scale=1.5] (i) at (18,7)   {%
    \begin{align*}
        \theta^\mu_{n+1}&=\begin{cases}
            \theta^\mu_{n}, &  L(\theta_n+\delta\theta)<L(\theta_n)\\
            \theta^\mu_{n}+\delta\theta^\mu, & L(\theta_n+\delta\theta)> L(\theta_n)
        \end{cases}\\
        \lambda_{n+1}&=\begin{cases}
            \lambda_{n}\lambda_u, & L(\theta_n+\delta\theta)< L(\theta_n)\\
            \frac{\lambda_{n}}{\lambda_d}, & L(\theta_n+\delta\theta)> L(\theta_n)
        \end{cases}
    \end{align*}
    };
    \node (j) at (0,-10){};
    \node (k) at (24,-10){};
    \draw[->,thick] (a)--(b);
    \draw[->,thick] (b)--(c);
    \draw[->,thick] (c)--(d);
    \draw[->,thick] (d)--(e);
    \draw[->,thick] (e)--(f);
    \draw[->,thick] (g)--(b);
    \draw[->,thick] (h)--(c);
    \draw[->,thick] (i)--(d);
  \draw[->,thick] (e)|-(j)-|(a);
\end{tikzpicture}  
\caption{Schematic overview of the weighted Levenberg–Marquardt (wLM) procedure. Blue callouts indicate steps common to both the standard and weighted LM algorithms; green callouts highlight steps specific to the wLM implementation.}\label{fig:Schema}
\end{figure*}

To allow for uncertainty in the relative reliability of each dataset, we introduce a weight parameter $\alpha_k$ for each group. These weights serve as inverse variance hyperparameters, allowing for potential under- or overestimation of the reported experimental uncertainties.

Due to the Central Limit Theorem \cite{feller1968}, we assume that the residuals are normally distributed. Under this assumption, the likelihood function for the data, conditioned on a set of dataset-specific weights  $\alpha = \{\alpha_1, \dots, \alpha_{N_g}\}$ takes the form \cite{Lahav2000}
\begin{align}
    L\left(y\mid\theta,\alpha\right)=\prod\limits_{k=1}^{N_g}\left(\frac{\alpha_k}{2\pi}\right)^{n_k/2}e^{-\frac{1}{2}\alpha_k \chi^2_k(\theta)}.
\end{align}
Following the procedure in \cite{Hobson2002}, we incorporate unknown weighting of the datasets in our optimization routine by marginalizing over the weighting parameters $\alpha_k$. 

The prior over each weight $\alpha_k$ is determined by maximizing the Shannon entropy \cite{Shannon1948,Jaynes1957}
\begin{align}
    S[p] = -\int_0^\infty p(\alpha_k) \ln p(\alpha_k)\, d\alpha_k,
\end{align}
subject to the constraints
\[
\int_0^\infty p(\alpha_k)\, d\alpha_k = 1 \quad \text{and} \quad \int_0^\infty \alpha_k p(\alpha_k)\, d\alpha_k = 1.
\]
This is a standard problem in the calculus of variations. Solving it using Lagrange multipliers yields the maximum entropy distribution:
\begin{align}
    p(\alpha_k) = e^{-\alpha_k},
\end{align}
i.e., the exponential distribution with unit mean. This choice reflects the assumption that the dataset is free from systematic biases, and that misquoted or unreported uncertainties can be accounted for through the weight parameter \( \alpha_k \).

Each dataset's likelihood contribution, $L(y_k \mid \theta, \alpha_k)$, is then integrated with respect to this prior to account for uncertainty in the error estimates. The marginalization over $\alpha_k$ yields the effective posterior likelihood:
\begin{align}
    L\left(\theta \mid y\right)&=\int\limits_{0}^\infty \mathrm{d}\alpha_1 e^{-\alpha_1}\cdots \int\limits_{0}^\infty \mathrm{d}\alpha_{N_g} e^{-\alpha_{N_g}}  L\left(y\mid\theta,\alpha\right)\\
    &=\prod\limits_{k=1}^{N_g}\frac{2\Gamma(n_k/2+2)}{\pi^{n_k/2}}
\left(\chi^2_k(\theta)+2\right)^{-(n_k+2)/2}.
\end{align}
This formulation implicitly weights each dataset based on its $\chi^2_k(\theta)$ value, modulating its relative contribution to the likelihood without the need for explicit assumptions about error inflation or deflation.

The exponential prior implicitly assumes an unbounded and rapidly decaying confidence in dataset reliability, which may not reflect prior knowledge or empirical uncertainty. To allow for more realistic and problem-specific modeling, we generalize this framework by replacing the exponential prior with an arbitrary distribution $ f(\alpha_k)$ defined on the interval $[\alpha_L, \alpha_U]$, thereby enabling greater flexibility in controlling the implicit dataset weighting.
The generalized likelihood function, $L$, is then
\begin{align}
    L\left(\theta \mid y\right)&=\int\limits_{\alpha_L}^{\alpha_U} \mathrm{d}\alpha_1  f(\alpha_1) \cdots \int\limits_{\alpha_L}^{\alpha_U}\mathrm{d}\alpha_{N_g} f(\alpha_{N_g})   L\left(y\mid\theta,\alpha\right)\\
   &=\prod\limits_{k=1}^{N_g}\frac{1}{(2\pi)^{n_k/2}} \mathcal{L}_{ \frac{\chi^2_k(\theta)}{2}}\left[\alpha_k^{n_k/2}  f(\alpha_k)\right],\end{align}
where $\mathcal{L}$ is the Laplace transform
\begin{equation}
    \mathcal{L}_s[f(x)]=\int\limits_0^\infty e^{-sx} f(x)\,\mathrm{d}x.
\end{equation}
The covariance matrix of model parameters can be estimated from the inverse of the Fisher information metric (FIM), denoted by $g_{\mu\nu}$, which is defined for any likelihood function as the negative expected value of the log-likelihood's Hessian:
\begin{align}
    g_{\mu\nu}&=-E\left[\partial_\mu\partial_\nu  \ln L\left(\theta\mid y\right)\right].
\end{align}
In the context of our weighted likelihood formulation, we define the \emph{wFIM} as $g^{(W)}_{\mu\nu}$. This matrix incorporates the influence of dataset-specific weights into the sensitivity structure of the likelihood. It takes the form:

\begin{align}
    g^{(W)}_{\mu\nu}&=-E\left[\partial_\mu\partial_\nu  \ln L\left(\theta\mid y\right)\right]\\
     &=\sum\limits_k g_{\mu\nu}^{(k)} w_k(\theta),
\end{align}
where $g_{\mu\nu}^{(k)}$ is the Fisher information contribution from the $k$-th dataset, and $ w_k(\theta)$ is the corresponding weighting function, defined as:

\begin{widetext}
   \begin{align}
   w_k(\theta)=\frac{ \mathcal{L}_{ \frac{\chi^2_k(\theta)}{2}}\left[\alpha_k^{n_k/2+1}  f(\alpha_k)\right]}{ \mathcal{L}_{ \frac{\chi^2_k(\theta)}{2}}\left[\alpha_k^{n_k/2}  f(\alpha_k)\right]}
    +\frac{\mathcal{L}_{ \frac{\chi^2_k(\theta)}{2}}\left[\alpha_k^{n_k/2+2}  f(\alpha_k)\right]}{ \mathcal{L}_{ \frac{\chi^2_k(\theta)}{2}}\left[\alpha_k^{n_k/2}  f(\alpha_k)\right]}
     -\frac{ \mathcal{L}^2_{ \frac{\chi^2_k(\theta)}{2}}\left[\alpha_k^{n_k/2+1}  f(\alpha_k)\right]}{ \mathcal{L}^2_{ \frac{\chi^2_k(\theta)}{2}}\left[\alpha_k^{n_k/2}  f(\alpha_k)\right]}.
\end{align}
\end{widetext}
We now consider several distinct functional forms for the prior distribution over the weighting variable $\alpha$, each of which leads to a different modification of the likelihood and ultimately each individual sub-data set's contribution to the cost function.

For the exponential distribution, i.e. $f(\alpha)=e^{-\alpha}$,
\newpage
\begin{widetext}
    \begin{align}
    g^{(W)}_{\mu\nu}&=E\left[\partial_\mu\partial_\nu  \ln L\left(\theta\mid y\right)\right]\\
    &= E\left[\sum\limits_{k=1}^{N_g}\frac{n_k+2}{2}\frac{(\chi^2_{k}+2)\partial_{\mu}\partial_{\nu}\chi^2_{k}-\partial_{\nu}\chi^2_{k}\partial_{\mu}\chi^2_{k}}{(\chi^2_{k}+2)^2}\right]\\
    &=\sum\limits_{k=1}^{N_g}\sum\limits_{i=1}^{n_k}(n_k+2)E\Big[\frac{\partial_\mu r^{(k)}{}^{i}(\theta) \partial_{\nu }r^{(k)}{}^{i}(\theta)}
        {(\chi^2_{k}+2)}-\frac{r^{(k)}{}^{i}(\theta)\partial_{\mu\nu}r^{(k)}{}^{i}(\theta)}
        {(\chi^2_{k}+2)}-2\sum\limits_{i=j}^{n_k}\frac{r^{(k)}{}^{i}(\theta)r^{(k)}{}^{j}(\theta)\partial_{\mu}r^{(k)}{}^{i}(\theta)\partial_{\nu}r^{(k)}{}^{j}(\theta)}{(\chi^2_{k}+2)^2}\Big],
\end{align}
\end{widetext}
where we used the shorthand notation for partial derivatives $\partial_\mu=\partial/\partial\theta^\mu$.
Since the measurements are independent, the expected value of the product of residuals is just $E\left[r^{(k)}{}^{i}(\theta)r^{(k)}{}^{j}(\theta)\right]=\delta^{ij}$. Therefore, we get the following expression
\begin{align}
    g^{(W)}_{\mu\nu}&=\sum\limits_{k=1}^{N_g}\sum\limits_{i=1}^{n_k}(n_k+2)\chi^2_k(\theta)\frac{\partial_{\mu}r^{(k)}{}^{i}(\theta)\partial_{\nu}r^{(k)}{}^{i}(\theta)}{(\chi^2_{k}+2)^2}\\
    &=\sum\limits_{k=1}^{N_g}w_k(\theta) g^{(0,k)}_{\mu\nu},
\end{align}
where we interpret this expression as a product of weights, $w_k$,
\begin{align}
   w_k(\theta) &=(n_k+2)\frac{\chi^2_k(\theta)}{(\chi^2_k(\theta)+2)^2}
\end{align}
 and the unweighted FIM for the subset $k$, $g^{(0,k)}_{\mu\nu}$,
\begin{equation}
    g^{(0,k)}_{\mu\nu}=\sum\limits_{i=1}^{n_k}\partial_{\mu}r^{(k)}{}^{i}(\theta)\partial_{\nu}r^{(k)}{}^{i}(\theta).
\end{equation}
 One can recover the standard, unweighted, FIM by setting $w_k(\theta)=1$
\begin{equation}
g^{(0)}_{\mu\nu}=\sum\limits_{k=1}^{N_g}g^{(0,k)}_{\mu\nu}.
\end{equation}
 
For the uniform distribution, the distribution of the weights is given by
\begin{equation}
    f(\alpha)=\frac{1}{\alpha_U-\alpha_L}\mathbb{1}_{\left[\alpha_L,\alpha_U\right]}(\alpha),
\end{equation}

while the required Laplace transform is
\begin{align}
    \mathcal{L}_{ \frac{\chi^2_k(\theta)}{2}}\left[\alpha_k^{n_k/2}  f(\alpha_k)\right]&=\frac{\gamma\left(\frac{n_k}{2},\alpha_U\frac{\chi^2_k(\theta)}{2}\right)-\gamma\left(\frac{n_k}{2},\alpha_L\frac{\chi^2_k(\theta)}{2}\right)}{(\alpha_U-\alpha_L)\left(\frac{\chi^2_k(\theta)}{2}\right)^{n_k/2+1}},
\end{align}
where $\gamma$ is the incomplete $\Gamma$ function
\begin{equation}
    \gamma(s,x)=\int\limits_0^x t^{s-1}e^{-t}\,\mathrm{d}t.
\end{equation}
We choose $\alpha_L$ and $\alpha_U$ so that the expected value of $\alpha$ 
\begin{align}
     E[\alpha]&=\int_{\alpha_L}^{\alpha_U} f(\alpha) \alpha\,\mathrm{d}\alpha = 1\\
    &=\frac{\alpha_L+\alpha_U}{2}.   
\end{align}
The easiest way to satisfy this condition is to set that the $\alpha_L$ and $\alpha_U$ are symmetrically positioned around unity on an interval of width $h$, i.e. $\alpha_L=1-h/2$, and $\alpha_U = 1+h/2$.

To interpolate between exponential decay and bounded support, we introduce a \emph{cropped exponential distribution}, which combines an exponential profile with compact support on an interval $[\alpha_L, \alpha_U]$. This form introduces a shape parameter $\eta$ that governs the decay rate:
\begin{equation}
    f(\alpha)=\frac{\eta\,e^{-\eta\alpha}}{e^{-\eta\alpha_L}-e^{-\eta\alpha_U}}\,\mathbb{1}_{[\alpha_L,\alpha_U]}(\alpha),
\end{equation}
where $\mathbb{1}_{[\alpha_L,\alpha_U]}(\alpha)$ is the indicator function restricting the support to $[\alpha_L, \alpha_U]$.

This distribution reduces to the standard exponential in the limit $\alpha_L = 0$, $\alpha_U \to \infty$. The expectation value is given by
\begin{align}
     E[\alpha]&=\int_{\alpha_L}^{\alpha_U} f(\alpha) \alpha\,\mathrm{d}\alpha = 1\\
    &=-e^{-\eta\alpha/2}\frac{e^{\eta\alpha}(\eta(\alpha-2)-2)+\eta(\alpha+2)+2}{4\eta\sinh\frac{\eta\alpha}{2}}.   \label{eq:croppedExponentialExpect}
\end{align}
To preserve the condition \( E[\alpha] = 1 \), we tune $\eta$ numerically or fix it and adjust the support accordingly.

The cropped-exponential distribution is also motivated by information-theoretic principles: it minimizes the Shannon entropy
\begin{align}
    S[p] = -\int_{\alpha_L}^{\alpha_U} p(\alpha) \ln p(\alpha)\, d\alpha,
\end{align}
subject to the constraints
\begin{equation}
\int_{\alpha_L}^{\alpha_U} p(\alpha)\, d\alpha = 1, \quad \int_{\alpha_L}^{\alpha_U} \alpha\, p(\alpha)\, d\alpha = 1.
\end{equation}
While inverting $E[\alpha] = 1$ to solve for $\eta$ can be numerically unstable in practice, we find satisfactory results by setting $\eta = 1$ and defining symmetric bounds $\alpha_L = 1 - h/2$ and $\alpha_U = 1 + h/2$, where $h$ controls the permitted range of variation.
The corresponding negative log-likelihood is 
\begin{equation}
    -\ln L= -\ln\left( \frac{\eta}{e^{-\eta \alpha_L}-e^{-\eta \alpha_U}}\frac{\Delta(n_k,\eta,\alpha_L,\alpha_U, \chi^2_k) }{(\chi^2_k/2+\eta)^{n_k/2+1}}\right),
\end{equation}
where we introduce the function 
\begin{align}
    \Delta(n_k,\eta,\alpha_L,\alpha_U, \chi^2_k) &= \gamma\left(n_k/2,\alpha_U\left(\chi^2_k/2+\eta\right) \right)\\
    &-\gamma\left(n_k/2,\alpha_L\left(\chi^2_k/2+\eta\right)\right).
\end{align}
The corresponding wFIM weights are
\begin{widetext}
    \begin{align}
    w_k=\frac{\Delta(n_k+2,\eta,\alpha_L,\alpha_U, \chi^2_k)}{\Delta(n_k,\eta,\alpha_L,\alpha_U, \chi^2_k)}
    +\frac{\Delta(n_k+4,\eta,\alpha_L,\alpha_U, \chi^2_k)}{\Delta(n_k,\eta,\alpha_L,\alpha_U, \chi^2_k)}
    -\left(\frac{\Delta(n_k+2,\eta,\alpha_L,\alpha_U, \chi^2_k)}{\Delta(n_k,\eta,\alpha_L,\alpha_U, \chi^2_k)}\right)^2.
\end{align}
\end{widetext}

In Fig.~\ref{fig:weights}, we illustrate the behavior of the log-likelihoods and the wFIM weights under the different weighting schemes discussed above. For this purpose, we consider a hypothetical dataset\footnote{A specific dataset is not required for this illustration---only the number of data points in each group matters, as the plotted quantities depend solely on $\chi^2_k$ and $n_k$.} comprising subsets of sizes $n_0 = 1$, $n_1 = 10$, and $n_2 = 100$, and evaluate the negative log-likelihoods and corresponding wFIM weights as functions of $\chi^2_k$. The uniform and cropped exponential weighting schemes are computed for three representative values of the width parameter $h$, while the exponential and unweighted schemes involve no additional parameters. The most permissive case, $h = 2$, corresponds to a lower bound $\alpha_L = 0$.

The negative log-likelihoods for large $\chi^2_k$ approach the unweighted line which is proportional to $\chi^2_k$, for all weightings. The wFIM weights for large $\chi^2_k$ and large $n_k$  behave similarly and approach unity. We can see that for small $\chi^2_k$ the weights behave differently for different priors. For the exponential prior the wFIM weights approach 0, while for the uniform prior they flatten-out at a value larger than 1. The cropped exponential weights remain close to unity for both large and small $\chi^2_k$.

For the nucleus examined in this study, the exponential weighting scheme consistently yielded lower overall $\chi^2$ values compared to the unweighted case and to both the uniform and cropped-exponential weighting strategies. However, this improvement in fit quality came at the cost of looser constraints on the parameter uncertainties. In a broader dataset encompassing a larger set of nuclei -- part of ongoing work -- we observed that the performance of each scheme varied
depending on both the amount and the consistency of
experimental data available for a given nucleus, indicating
that the most effective weighting choice is inherently
data-dependent. We find in practice that for this wider set of nuclei we can achieve convergence with $h=0.4$. Larger values of $h$ yielded equivalent results to smaller $h$ for some nuclei, but gave unstable fits for others. 
\subsection{Weighted Levenberg-Marquardt algorithm}
The classical Levenberg-Marquardt method \cite{marquardt1963algorithm} is a minimization procedure based on the derivatives of the likelihood function, incorporating Tikhonov regularization as a damping parameter. Briefly, this method leverages the residuals, $r^i$, between the data and the model, together with the Jacobians of the residuals, $\partial_\mu r^i$, to compute an update step in the parameter space:
\begin{equation}
    \delta \theta^\mu = -\left[(g + \lambda D)^{-1}\right]^{\mu\nu} \sum_{k=1}^{N_g} \sum_{i_k=1}^{n_k} \partial_\nu r^{i_k} \, r^{i_k},  \label{eq:thetaupdate}
\end{equation}
where $g_{\mu\nu}$ is the Fisher information metric (FIM),
\begin{equation}
    g_{\mu\nu} = \sum_{k=1}^{N_g} \sum_{i_k=1}^{n_k} \partial_\mu r^{i_k} \, \partial_\nu r^{i_k},
\end{equation}
and $D_{\mu\nu}$ is a chosen damping matrix.

At each iteration, the FIM is computed and combined with a damping term $\lambda$ to ensure numerical stability and convergence of the minimization step. The parameter covariance matrix, $[\Sigma_\theta]^{\mu\nu}$, can be estimated from the Cramér-Rao bound as the inverse of the FIM \cite[see, e.g.,][]{Cramer99}.

\begin{figure*}
 \centering\
    \includegraphics[width=.8\linewidth]{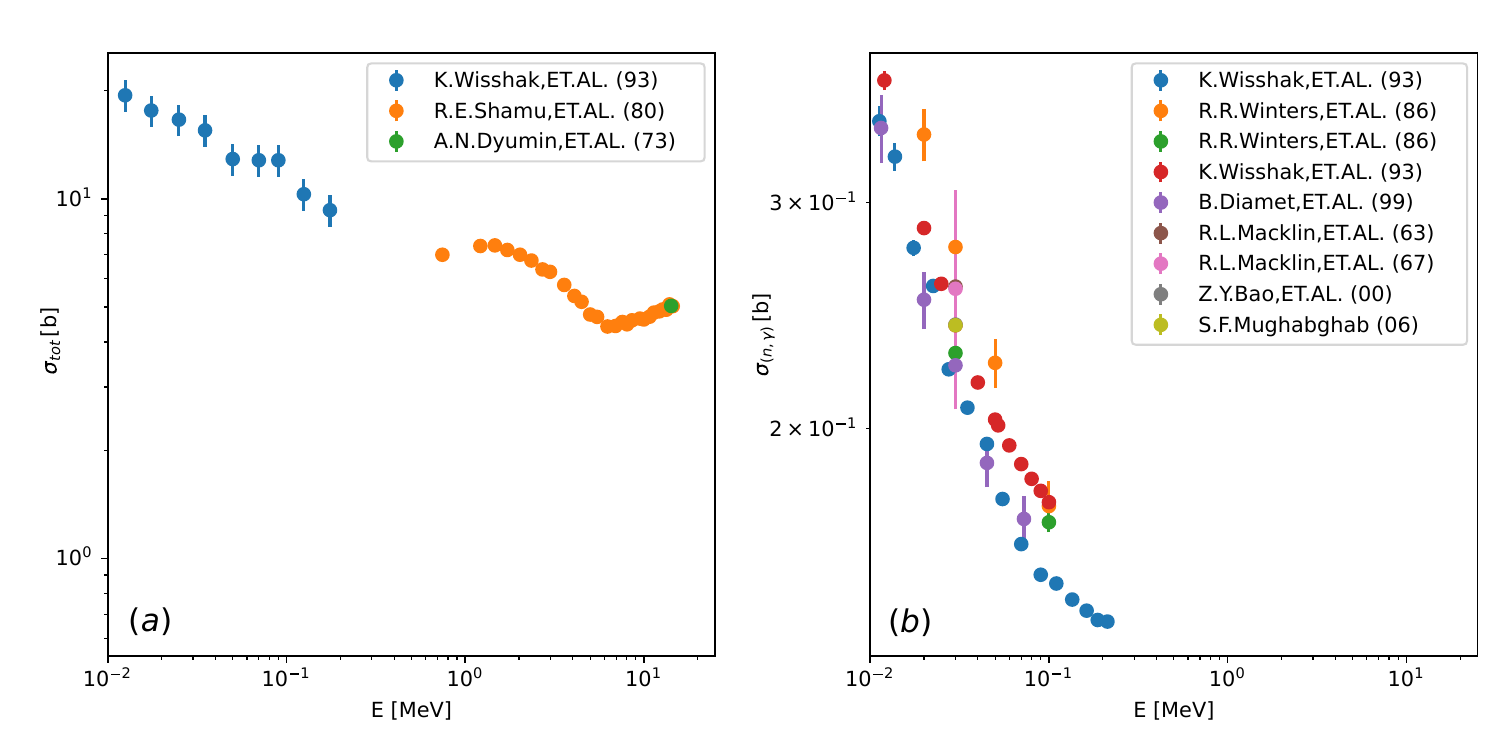}\
    \caption{Experimental data for (a) total and (b) capture cross section data for neutrons on $^{148}$Sm.}
    \label{fig:DataSm148}
    \centering
    \includegraphics[width=.8\linewidth]{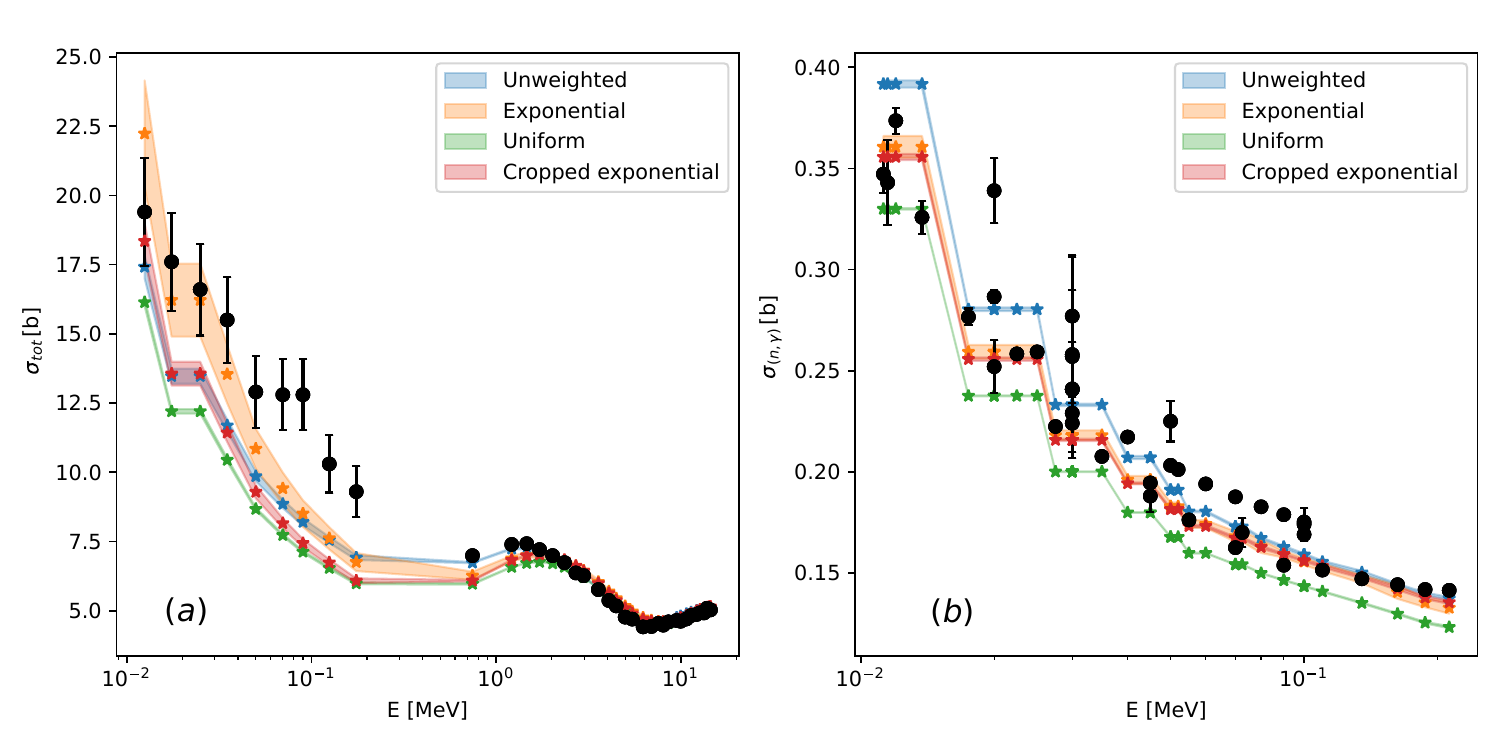}
   \caption{Best-fitting CoH$_3$ model predictions for (a) total and (b) capture cross sections for neutrons on $^{148}$Sm, using unsmoothed data. Shown are the fitted curves and their $1\sigma$ confidence intervals under various weighting schemes. Experimental data from Fig.~\ref{fig:DataSm148} are shown in black.}
    \label{fig:EvalsSm148}
    \begin{subfigure}[b]{0.5\textwidth}
        \includegraphics[width=\linewidth]{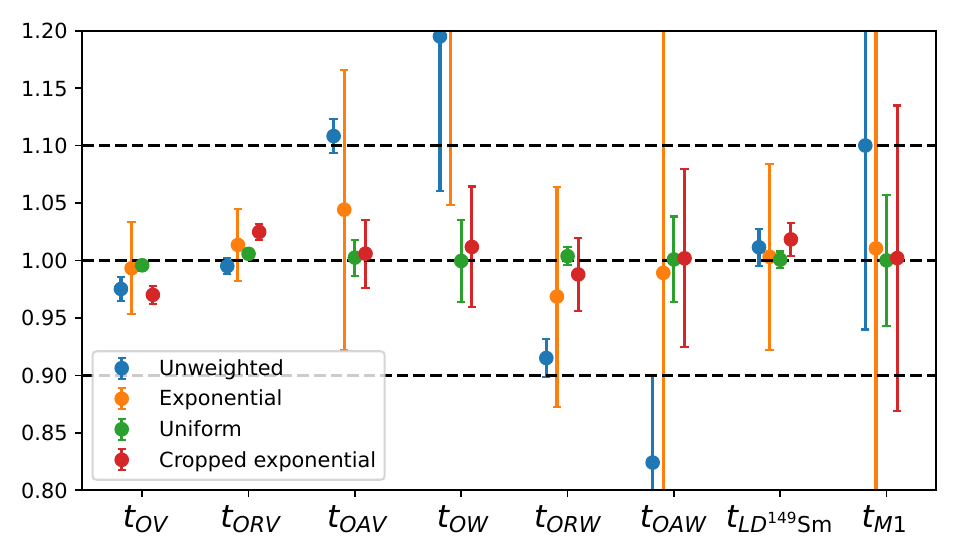}\
        \end{subfigure}\begin{subfigure}[b]{0.5\textwidth}    
     \includegraphics[width=\linewidth]{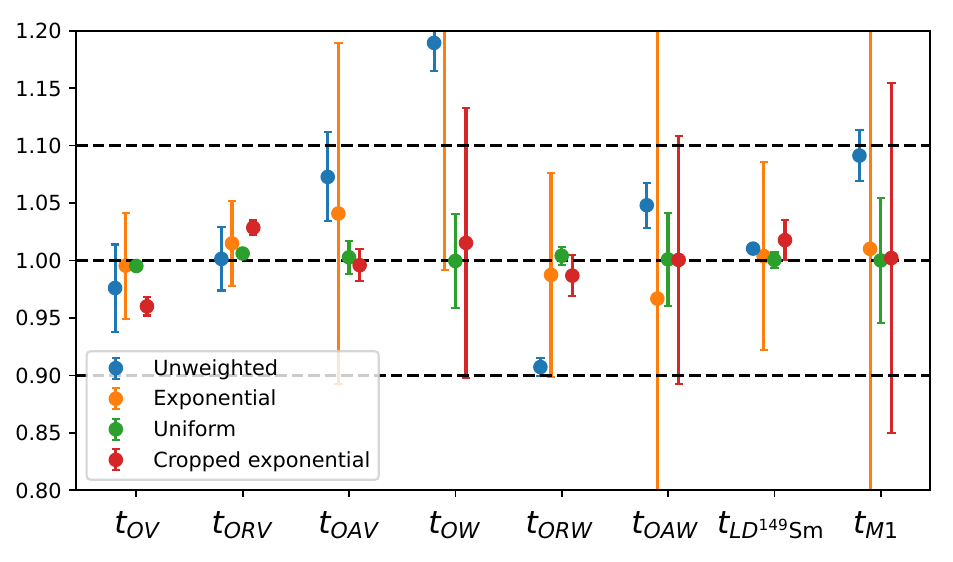}
    \end{subfigure}
    \caption{Best-fit CoH$_3$ model parameters for the fits shown in Fig. \ref{fig:EvalsSm148} with $1\sigma$ confidence intervals. (a) Results without geometric scaling of the update step in the wLM procedure and (b) results including geometric scaling. Dashed black lines indicate a reasonable/physical range of the parameters.}
    \label{fig:FitsSm148}
\end{figure*}

\begin{figure*}\
    \centering\vspace{-.1cm}
    \includegraphics[width=.8\linewidth]{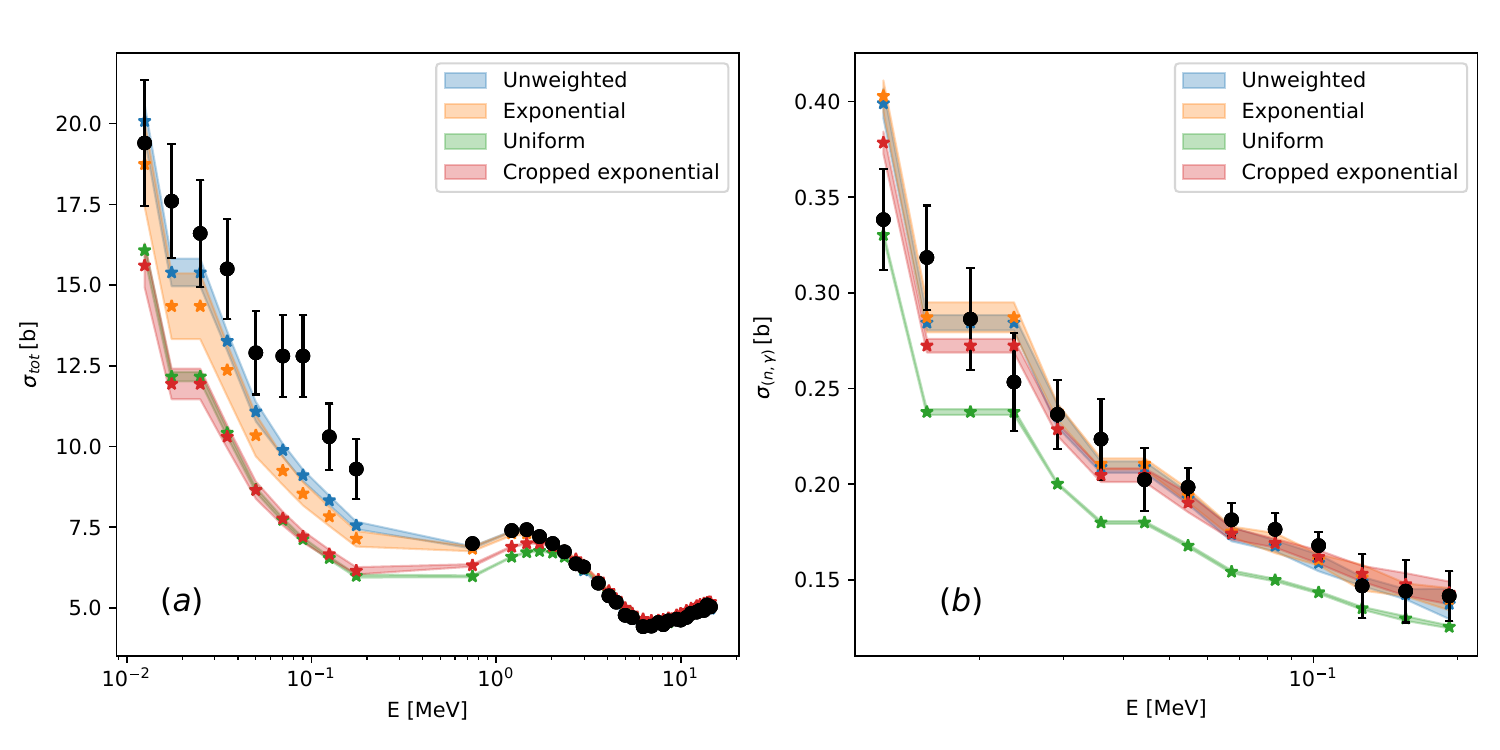}\
    \caption{Same as Fig. \ref{fig:EvalsSm148} where FBET smoothing was used for the capture data. Black curves show the unsmoothed total cross section from Fig.~\ref{fig:DataSm148}. The capture channel data has been smoothed via FBET; error bars reflect the diagonal of the resulting covariance matrix.}\
    \label{fig:EvalsSm148Smooth}\
    \begin{subfigure}[b]{0.5\textwidth}\    
        \includegraphics[width=\linewidth]{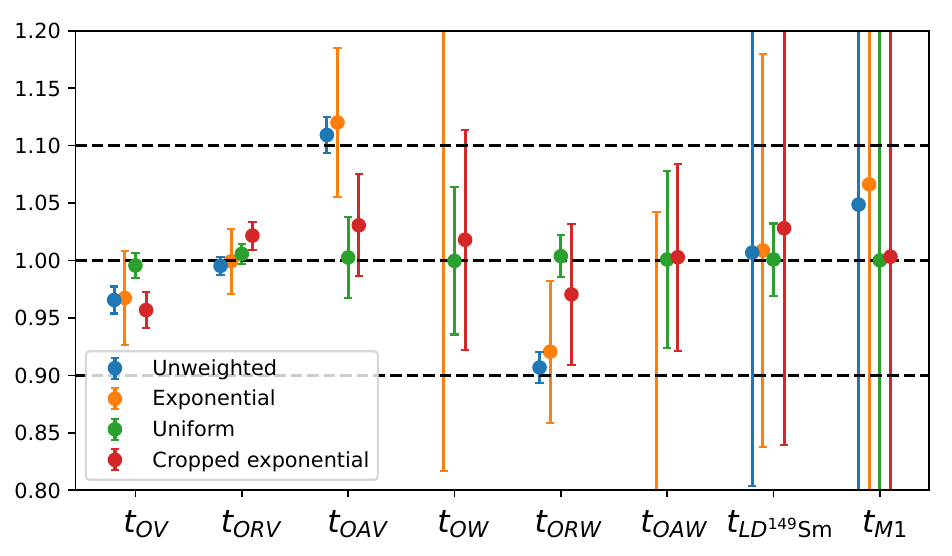}\
        \end{subfigure}\begin{subfigure}[b]{0.5\textwidth}\    
     \includegraphics[width=\linewidth]{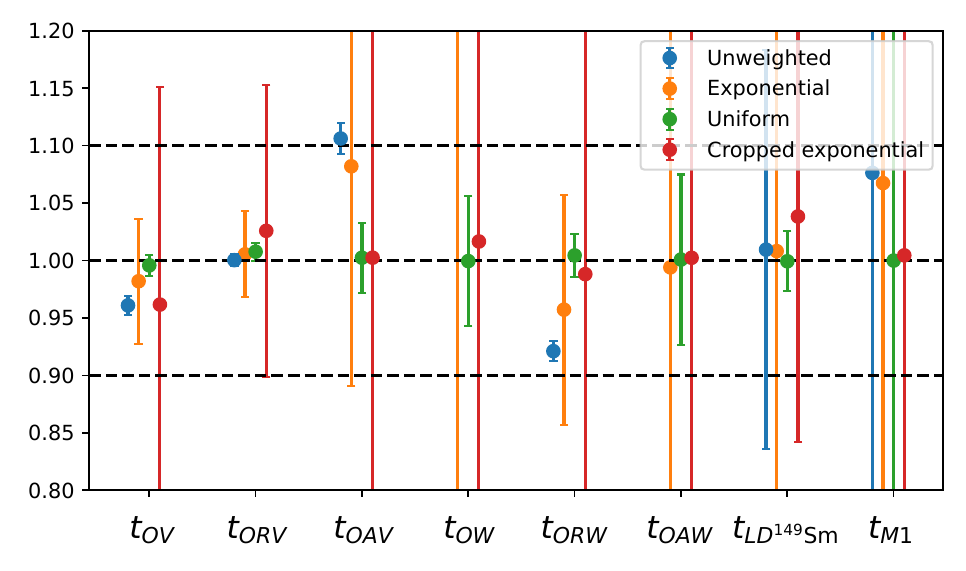}
    \end{subfigure}
    \caption{Best-fit CoH$_3$ model parameters for total and smoothed capture channel cross section fits to $^{148}$Sm, with $1\sigma$ confidence intervals.     
    The unweighted (classic) fit yields parameter values for the imaginary surface potential depth, $t_{OW}$, and the imaginary surface potential diffuseness, $t_{OAW}$, that lie outside the expected physical range. Left: without geometric scaling. Right: with geometric scaling of the update step in wLM.}
    \label{fig:FitsSm148Smooth}\vspace{-.1cm}
      \centering\
    \includegraphics[width=.5\linewidth]{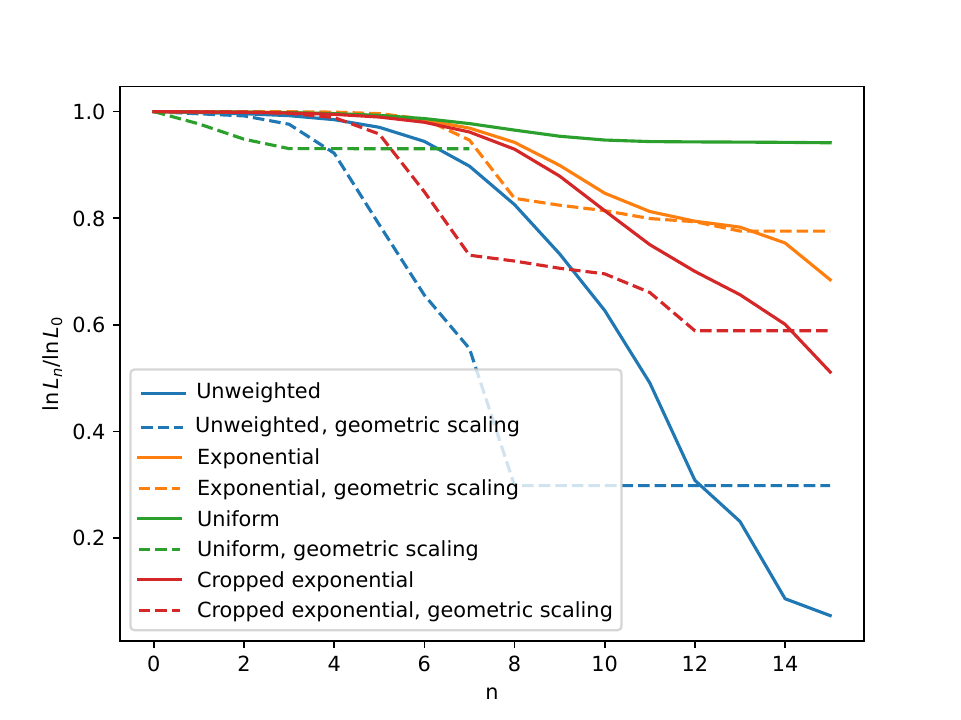}\
    \caption{Log-likelihood ratio evolution over the number of optimization steps, $n$, for each weighting scheme. Solid lines show standard wLM optimization; dashed lines include geometric scaling of the $\delta\theta$ step.}\
    \label{fig:modifiedSteps}\
\end{figure*}
\begin{figure*}

    \includegraphics[width=\linewidth]{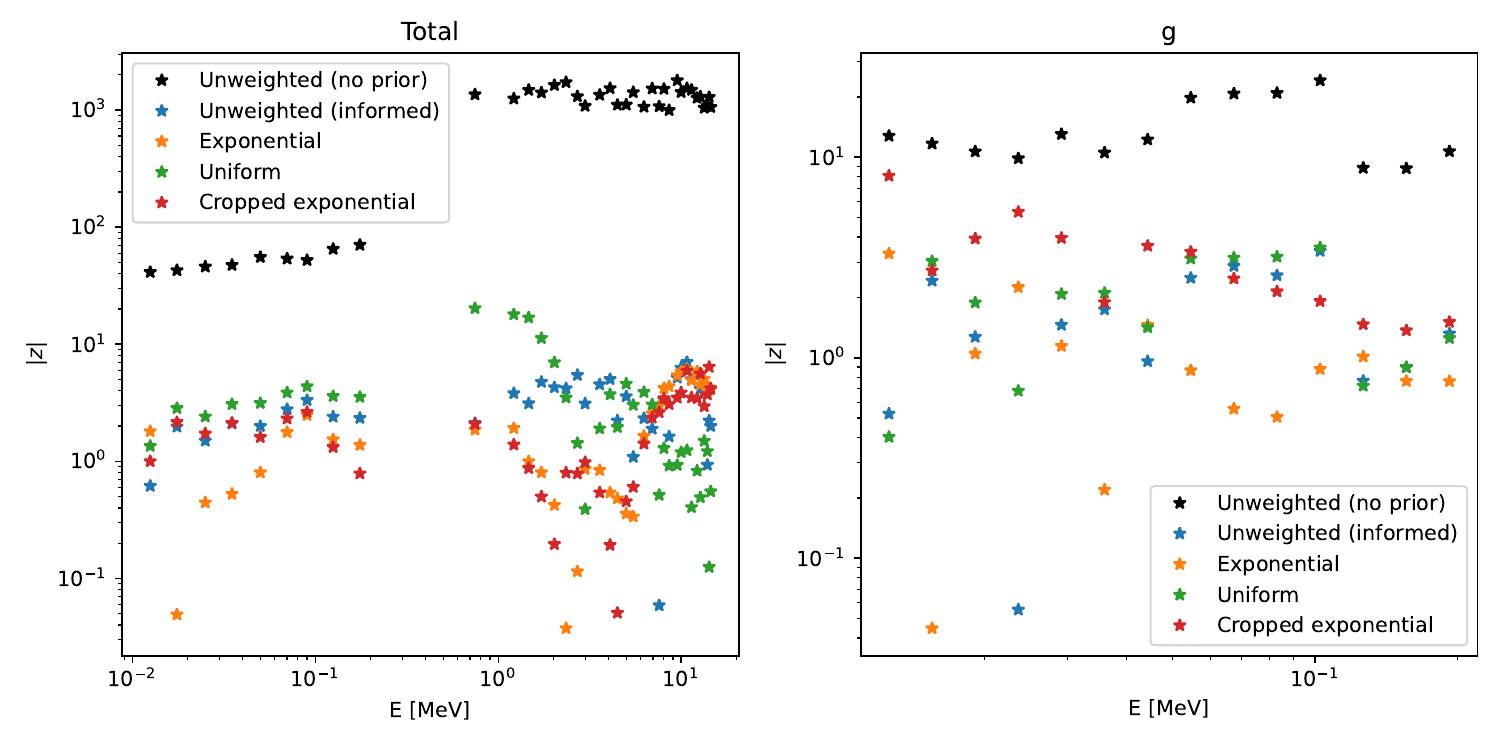}
    \caption{Absolute z-scores for various fitting schemes, evaluated against $^{148}$Sm cross section data for the (a) total and (b) capture channels. Included are the unweighted LM (black), LM initialized at the best-fit exponential wLM parameters (blue), and the exponential (orange), uniform (green), and cropped-exponential (red) wLM fits.}\label{fig:zscores}
\end{figure*}
\begin{figure}
    \centering    \includegraphics[width=\linewidth]{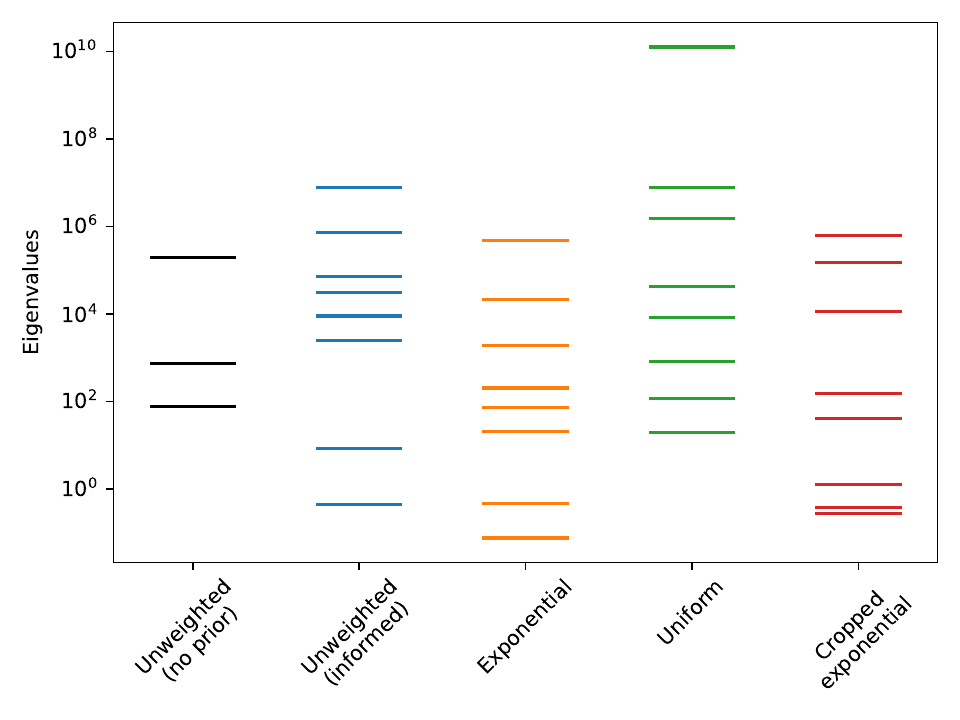}
    \caption{(w)FIM eigenvalue spectra for various minimization strategies applied to $^{148}$Sm. From left to right: classical unweighted LM (black), LM initialized at the exponential wLM optimum (blue), exponential wLM (orange), uniform wLM (green), and cropped-exponential wLM (red).}
    \label{fig:spectrumWLM}
\end{figure}
We can use the wFIM derived above to derive the wLM algorithm by using the weights $w_k$ of the different datasets 

 \begin{equation}
     \delta\theta^\mu = -g^{(W,\lambda)}{}^{\mu\nu}\sum\limits_{k=1}^{N_g} \sum\limits_{i_k=1}^{n_k}w_k(\theta) \partial_\nu r^{i_k} r^{i_k},  \label{eq:thetaupdate2}
 \end{equation}
 where we have applied L2 regularization to the wFIM that is also used for the standard LM algorithm \citep[see, e.g., ][]{Transtrum2012}
 \begin{equation}
     g^{(W,\lambda)}_{\mu\nu}=g^{(W)}_{\mu\nu} + \lambda D_{\mu\nu},
 \end{equation}
 where $D_{\mu\nu}$ is the diagonal matrix
 \begin{equation}
     D_{\mu\nu}=\begin{cases}
          g^{(W)}_{\mu\nu}, & \mu=\nu\\
          0, & \mu\neq \nu
     \end{cases}.
 \end{equation}
 For a brief discussion on the impact of $\lambda$, see appendix \ref{sec:damping}.
 The FIM with upper $\mu,\nu$ indices in Eq. \ref{eq:thetaupdate2} is the matrix inverse of the FIM, in the sense that \begin{equation}g^{(W,\lambda)}_{\mu\nu} g^{(W,\lambda)}{}^{\nu\rho}=\delta_\mu^\rho.\end{equation}
 
 We implement the change of $\lambda$ analogously to the standard LM procedure. In Fig. \ref{fig:Schema}, we summarize the steps of the wLM procedure, where we have highlighted in green where the weighting is included compared to the original algorithm.  The proposal in the $n$-th step is given recursively, based on the total sum of residuals criterion:
 \begin{align}
\theta^{\mu}_{n+1}&=\begin{cases}\theta^{\mu}_{n-1}, & -\ln L(\theta_{n-1}+\delta\theta)> -\ln L(\theta_{n-1})\\
     \theta^{\mu}_{n-1}+\delta\theta^\mu, & -\ln L(\theta_{n-1}+\delta\theta)< -\ln L(\theta_{n-1})
        \end{cases}\\
        \lambda_{n+1}&=\begin{cases}
            \lambda_{n-1} \lambda_u, & -\ln L(\theta_{n-1}+\delta\theta)> -\ln L(\theta_{n-1})\\
     \frac{\lambda_{n-1}}{\lambda_d}, & -\ln L(\theta_{n-1}+\delta\theta)< -\ln L(\theta_{n-1})
        \end{cases},
 \end{align}
 where we have achieved satisfactory results with the standard choice of the hyperparameters $(\lambda_u,\lambda_d)=(10,2)$. We start the minimization algorithm at the initial step $\theta^\mu_0=1$, $\lambda_0=10^3$.
After $N$ minimization steps, we estimate the parameter covariance matrix using the expression as the inverse of the wFIM.
\subsection{Geometric Scaling – Adaptive Step Control via Angular Alignment}

In optimization landscapes with strong curvature---such as those encountered in \textit{sloppy} models---the parameter space often contains narrow, winding valleys where optimal solutions reside. Fixed step-size descent methods are prone to overshooting or becoming unstable in such regions, especially when entering areas of rapidly changing curvature.

Geometric scaling introduces an adaptive mechanism based on the \emph{angle between successive parameter updates}, $\delta\theta_n$ and $\delta\theta_{n-1}$, evaluated using the weighted Fisher Information Metric (wFIM), to modify the wLM update by a scaling factor, $s$,
\begin{align}
s &= g^{(W,\lambda)}_{\mu\nu} \, \delta\theta^\mu_{n-1} \, \delta\theta^\nu_n\\
\delta\theta^\alpha_n &\to s \delta\theta^\alpha_n.\end{align}

If the angle is small (updates are aligned), the optimization is likely following a stable descent direction within a curved valley --- the step can be extended. If the angle is large (abrupt directional change), the trajectory is likely hitting curvature boundaries --- the step should be scaled down.
This method acts as a deterministic counterpart to phenomena commonly exploited in machine learning optimizers, such as combinations of momentum \cite{Qian1999,Neal2011}, stochastic gradients \cite{Feng2021}, and decay\cite{Loschilov2017}. Here, however, the geometric structure of the model itself dictates the step adaptation.

As demonstrated in Sect.~\ref{sec:results}, the scaling mechanism \emph{does not affect} the final converged parameter estimates, but it significantly accelerates convergence---especially in high-dimensional or ill-conditioned parameter spaces. We show that this geometric scaling improves convergence rates consistently, regardless of whether the weighted or unweighted approach is used.
\section{Methods}\label{sec:methods}
  In sections \ref{sec:fbet} and \ref{sec:coh} we briefly describe the standard FBET techniques and the CoH$_3$ optical potential model.
\subsection{FBET}\label{sec:fbet} 
Prior information is often incorporated through a linearized approximation of Bayes' rule \cite[refer to][for details]{leeb_consistent_2008}. Experimental nuclear cross sections $\mathbf{y}$ evaluated from measurements $\mathbf{y}_m$ at chosen grid points using a function $f_M(\mathbf{y}_m)$ for an evaluation model $M$.  The linearization process yields estimates for the mean posterior value, $\mathbf{y}_1$, along with the corresponding posterior covariance matrix $\Sigma$. This approach presumes that the priors follow a normal distribution:

\begin{equation} p(\mathbf{y} \mid M) = N \exp\left(-\frac{1}{2} (\mathbf{y} - \mathbf{y}_0)^T A_0^{-1}(\mathbf{y} - \mathbf{y}_0)\right), \end{equation} where $\mathbf{y}_0$ represents the prior mean value of the parameter vector $\mathbf{y}$, and $A_0$ is the prior covariance matrix. The normalization constant $N$ (and $N',$ introduced later) ensures that $p(\mathbf{y} \mid M)$ integrates to unity over all possible $\mathbf{y}$; its exact value is irrelevant for the discussion that follows.

Assuming that the experimental data are associated with uncertainties characterized by a covariance matrix $B$, the posterior distribution is then given by:

\begin{align} p(\mathbf{y} \mid M) &= N' \exp\Big(-\frac{1}{2} (\mathbf{y} - \mathbf{y}_0)^T A_0^{-1}(\mathbf{y} - \mathbf{y}_0)\\
&- \frac{1}{2} (f_M(\mathbf{y}) - \mathbf{y}_m)^T B^{-1} (f_M(\mathbf{y}) - \mathbf{y}_m)\Big), \end{align}
were $N'$ is a new normalization constant that absorbs the contributions from both the prior and likelihood terms.

 The sensitivity matrix is introduced to linearize the relationship between $y_1$ and $f(\mathbf{y}_m)$. Following \cite{schnabel_fitting_2018}, we construct the matrix $S$, which effectively evaluates the measurements on grid points, using the prototypic model. In our case of nuclear cross sections, $\mathbf{y}_m=(y_1,\cdots, y_n)$ values represent the cross sections with their respective errors $\mathbf{\sigma}_m=(\sigma_1,\cdots,\sigma_n)$, while the $\mathbf{x}_m=(x_1,\cdots,x_n)$ values represent the energies at which the cross sections were measured. 
The function $f$ is evaluated on arbitrary grid points $\mathbf{x}$
\begin{equation}
    f_M(\mathbf{y})=\frac{\sum\limits_i y_i \phi\left(\mathbf{x}\mid x_i,\epsilon\right) }{\sum\limits_i \phi\left(\mathbf{x}\mid x_i,\epsilon\right)}=S^T\mathbf{y},
\end{equation}
where $\phi$ are the Gaussians \begin{equation}
    \phi\left(x\mid x_i,\epsilon\right)=\frac{e^{-\frac{(x-x_i)^2}{2\epsilon^2}}}{\sqrt{2\pi}\epsilon}.
\end{equation}
The matrix $B$ can simply be taken as the diagonal covariance matrix of measurement errors, $B=\mathrm{diag}(\mathbf{\sigma}^2_m)$ \footnote{Informally, and by stepping outside Einstein summation convention, one might write this as $B_{ii}=[\sigma^2_m]_i$}. 
Consequently, this procedure provides linearized estimates for $\mathbf{y}_1$ and $\Sigma$:

\begin{align} \mathbf{y}_1 &= \mathbf{y}_0 + A_0 S^T (S A_0 S^T + B)^{-1} (\mathbf{y}_m -S\mathbf{y}_0),\\ \Sigma &= A_0 - A_0 S^T (S A_0 S^T + B)^{-1} S A_0. \end{align}

We model $A_0$ to be proportional to a matrix described by a correlation factor $\rho$. 
\begin{equation}
    A_0 = \mathbf{s} \mathbf{s}^T(\mathbb{1}+\rho (\mathbf{1}\mathbf{1}^T-\mathbf{1})).
\end{equation}
We take factors $\epsilon$ and $\mathbf{s}$ to be functions of the measurements' energies and gridpoint energies, $\epsilon_i= \epsilon^0+\epsilon^1 x_i$ and $\mathbf{s}(\mathbf{x})=s^0+s^1 \mathbf{x}$, respectively. We perform a grid search to find the coefficients $\epsilon^0, \epsilon^1, s^0, s^1, \rho$ that minimize the generalized $\chi^2$ value \cite{schnabel_fitting_2018} 
\begin{equation}
   \chi^2\begin{pmatrix}\epsilon^0\\ \epsilon^1\\ s^0\\ s^1\\ \rho \end{pmatrix}=(\mathbf{y}_m-S\mathbf{y}_0)^T (S A_0 S^T+B)^{-1}(\mathbf{y}_m-S\mathbf{y}_0).
\end{equation}
\subsection{CoH$_3$ optical model fitting}
\label{sec:coh}
The CoH$_3$ code provides comprehensive calculations of nuclear reaction cross sections in the fast-neutron region, above the resonance range. It includes contributions from total and shape elastic cross sections, direct inelastic and direct/semidirect capture, pre-equilibrium emission, and both particle and $\gamma$-ray emission channels \cite{Kawano2019}. The code internally computes particle transmission coefficients, eliminating the need for external optical model solvers.

To describe reactions on deformed nuclei, CoH$_3$ employs coupled-channels calculations augmented by rotational and vibrational collective models. In our study, we focused on the neutron Koning--Delaroche optical potential  \citep{Koning2003}, modified by multiplicative scaling factors, or `tweaks' in CoH$_3$ terminology, of the real potential depth, $t_{OV}$, the imaginary surface potential depth, $t_{OW}$, the real potential radius, $t_{ORV}$, the imaginary surface potential radius, $t_{ORW}$, the real potential diffuseness, $t_{OAV}$, and the imaginary surface potential diffuseness, $t_{OAW}$.
Additionally, we considered tweaks to the M1 strength function, $t_{M1}$, and to the level density of the compound nucleus $t_{LD{}^{A}X}$. These parameters were fitted to experimental total and capture cross section data for the n+$A$ reaction for incident neutron energies ranging from roughly 0.01~MeV to 10~MeV.

\section{Results}\label{sec:results}
We demonstrate our minimization techniques on nuclear cross section data for both the total \cite{Wisshak1993,Shamu1980,Dyumin1973} and capture \cite{Wisshak1993,Winters1986,Duamet1999,Macklin1963,Macklin1967,Bao2000,Mughabghab2006} reactions for neutrons on ${}^{148}$Sm, as plotted in Fig.~\ref{fig:DataSm148}, were obtained from the EXFOR database \cite{EXFOR2018}. The n+${}^{148}$Sm reaction serves as a representative case from a broader dataset to be analyzed in a forthcoming publication. This nucleus was selected due to the availability of multiple experimental datasets per reaction channel and the relatively smooth behavior of its cross sections. There seems to be no resonance structure over the measured energy range that could be seen in the experimental data. These features make it particularly well-suited for demonstrating the wLM algorithm, in contrast to more resonance-dominated cases where there is a need for more data pre-processing techniques, even prior to FBET \cite{Imbrisak2025c}. We applied the wLM method both on raw experimental data and data that had been FBET smoothed.

Figure~\ref{fig:EvalsSm148} shows the outcomes of the various minimization schemes with different data weighting. The unweighted (classic, blue) fit tends to overfit one channel at the expense of the other, while the weighted schemes (orange, green, red) more effectively balance agreement across both. This advantage of weighting becomes especially apparent in Fig.~\ref{fig:FitsSm148}, where the unweighted fit parameters lie closer to the edges of the acceptable parameter region. For this dataset, the exponential weighting scheme yielded the best alignment with the experimental data.

By tempering the influence of overrepresented or overly precise data groups, each weighting scheme regularizes the contribution of groups with extreme likelihoods. This results in fit parameters that remain consistent with theoretical expectations and within feasible bounds---thereby improving interpretability and generalizability, especially in high-dimensional or underdetermined settings.

We also tested our method on a dataset in which the experimental data in the capture channel was smoothed using FBET. As shown in Fig.\ref{fig:EvalsSm148Smooth}, the cropped exponential weighting scheme outperformed the other approaches, indicating that the optimal choice of weighting prior may be dataset-dependent. The best-fit parameters obtained (Fig.\ref{fig:FitsSm148Smooth}) are consistent with those from the unsmoothed case, supporting the robustness of the method even when smoothing is applied. Smoothing becomes necessary when working with larger datasets, as it reduces the number of required model evaluations.

Furthermore, Fig.~\ref{fig:modifiedSteps} illustrates the benefit of geometric scaling during optimization as it enables more efficient improvement of the log-likelihood function across iteration steps. While the best-fit parameters obtained with and without geometric scaling (shown in the right and left panels of Figs.~\ref{fig:FitsSm148} and \ref{fig:FitsSm148Smooth}, respectively) are similar, geometric scaling significantly reduces the number of wLM iterations needed to reach convergence. By aligning step sizes with the local geometry, geometric scaling acts similarly to a preconditioner, since different parameter directions have different local sensitivities or curvature scales.

To further analyze the fits, Fig.~\ref{fig:zscores} displays the absolute z-scores, i.e., the difference between cross sections evaluated for each scheme and the smoothed experimental data normalized by measurement uncertainties. These provide an alternative, more granular assessment of performance than the total $\chi^2$. The wLM algorithms treated low- and high-energy measurements on more equal footing, despite their differing uncertainties and experimental sources. Interestingly, the classical LM algorithm disproportionately emphasized high-energy data, due to their smaller reported uncertainties, but with considerably larger z-scores. When the classical LM procedure was initialized close to the best-fitting value obtained via exponential wLM, the absolute z-scores were reduced to the same order of magnitude as the weighted fits. This again highlights the importance of informed initialization in high-dimensional, sloppy models. While the uniform wLM performed comparably to exponential wLM across much of the energy range, it was heavily penalized for deviations near $\sim$1 MeV.

The structural differences among the optimization schemes are also evident in the FIM spectra shown in Fig.~\ref{fig:spectrumWLM}. All schemes reveal the inherent sloppiness of the CoH$_3$ model, with FIM eigenvalues spanning many orders of magnitude. The classical LM algorithm’s FIM was nearly degenerate, with only three nonzero eigenvalues---explaining its failure to converge unless properly initialized. 

While the exponential scheme optimizes for $\chi^2$, it produces fits with larger uncertainties, as reflected in the smallest spectrum of FIM, and therefore through Cramér-Rao bound, high dispersion of posterior parameter estimates. Interestingly, this effect appears consistently across both raw and FBET-smoothed datasets, indicating that the tradeoff is not an artifact of data pre-processing but a universal feature of the weighting formulation. From an information-geometric perspective, this highlights a deeper structural insight: when comparing weighting schemes a lower $\chi^2$ does not imply higher FIM eigenvalues. In fact, the opposite may occur when the model is allowed to concentrate its degrees of freedom in narrow, high-confidence regions of data space, such as the $>1\,\mathrm{MeV}$ region for the total cross section in our case.

These findings suggest that modelers should compare $\chi^2$-optimal solutions obtained by different dataset weightings with caution, particularly when such solutions lie near or beyond physically plausible parameter boundaries and exhibit large posterior uncertainties. In high-dimensional, sloppy models, the \emph{best} fit may, counterintuitively, be the one with the \emph{largest} uncertainties. 
\section{Conclusion\label{sec:conclusion}}

The standard Levenberg–Marquardt algorithm, while widely used for nonlinear least squares minimization, encounters significant challenges in high-dimensional, ill-conditioned models such as CoH$_3$. These “sloppy” models exhibit vast variations in parameter sensitivities, often causing slow convergence or entrapment near non-physical boundaries in parameter space.

In this work, we introduced a weighted extension of the Levenberg–Marquardt algorithm (wLM) that incorporates dataset-specific weights through a generalized likelihood framework. This enables the construction of a weighted Fisher Information Metric (wFIM) which balances heterogeneous datasets and naturally guides optimization toward physically meaningful regions---eliminating the need for hard parameter boundaries. We demonstrated that flexible prior distributions over these weights can be tailored to improve convergence across diverse dataset structures, including both raw and FBET-smoothed nuclear cross section data.

We also proposed a geometric scaling technique based on the angular alignment of successive optimization steps. This deterministic strategy enhances optimization efficiency without sacrificing fit accuracy, providing a principled way to navigate the curved likelihood manifold inherent to complex models.

Our results using the example of neutron-induced reactions on ${}^{148}$Sm showcase the robustness and versatility of this framework. Aside from nuclear reactions, this methodology holds promise as a foundation for model calibration and constrained optimization in other scientific domains characterized by data heterogeneity and parameter degeneracy. We envision it enabling more reliable, physically consistent parameter estimation in increasingly complex modeling challenges ahead.
\appendix
\section{Impact of damping}\label{sec:damping}
The weighting factor $\lambda$ introduces a bias in Eq.~\ref{eq:thetaupdate2}. For small values of $\lambda$, we perform a Taylor series expansion near $\lambda=0$. This yields a small correction relative to the step computed in the absence of damping, denoted as $\delta\theta^{\mu}(\lambda=0)$:
\begin{widetext}
\begin{align}
     \delta\theta^\mu &= -[(g + \lambda D)^{-1}]^{\mu\nu}\sum\limits_{k=1}^{N_g} \sum\limits_{i_k=1}^{n_k}w_k(\theta) \partial_\nu r^{i_k} r^{i_k}\\
     &=-\left[g^{-1} -\lambda g^{-1}D g^{-1}+O(\lambda^2)\right]^{\mu\nu}\sum\limits_{k=1}^{N_g} \sum\limits_{i_k=1}^{n_k}w_k(\theta) \partial_\nu r^{i_k} r^{i_k}.
     \end{align}
    For simplicity, we define $\delta\theta^{\mu}(\lambda=0)$ as the step computed at $\lambda=0$, so that the expansion becomes
     \begin{align}
    \delta\theta^\mu  &=-\left(\delta^\mu_\beta-\lambda g^{\mu\alpha}D_{\alpha\beta}\right)g^{\beta\nu}\sum\limits_{k=1}^{N_g} \sum\limits_{i_k=1}^{n_k}w_k(\theta) \partial_\nu r^{i_k} r^{i_k}+O(\lambda^2)\\
     &=\delta\theta^{\mu}(\lambda=0)-\lambda g^{\mu\alpha}D_{\alpha\beta}\delta\theta^{\beta}(\lambda=0)+O(\lambda^2).
 \end{align}

 The limit $\lambda\to\infty$ can be derived assuming that wFIM has no zeros on its diagonal, $D$. 

 \begin{align}
     \delta\theta^\mu(\lambda\to\infty) &= -\frac{1}{\lambda}\left[\left(D+\frac{1}{\lambda} g \right)^{-1}\right]^{\mu\nu}\sum\limits_{k=1}^{N_g} \sum\limits_{i_k=1}^{n_k}w_k(\theta) \partial_\nu r^{i_k} r^{i_k}\\
     &=-\frac{1}{\lambda}\left[D^{-1} -\frac{1}{\lambda} D^{-1}g D^{-1}+O(\lambda^{-2})\right]^{\mu\nu}\sum\limits_{k=1}^{N_g} \sum\limits_{i_k=1}^{n_k}w_k(\theta) \partial_\nu r^{i_k} r^{i_k}\\
     &=-\frac{1}{\lambda}\left(D^{\mu\nu}-\frac{1}{\lambda} D^{\mu\alpha}g_{\alpha\beta}D^{\beta\nu}\right)\sum\limits_{k=1}^{N_g} \sum\limits_{i_k=1}^{n_k}w_k(\theta) \partial_\nu r^{i_k} r^{i_k}+O(\lambda^{-3}).
     \end{align}

     We insert the expression for $\delta\theta^{\sigma}(\lambda=0)$ and compute the first-order behavior of $\delta\theta^{\sigma}(\lambda\to\infty)$
     
         \begin{align}
  \delta\theta^\mu (\lambda\to\infty)&=-\frac{1}{\lambda}\left(D^{\mu\gamma}-\frac{1}{\lambda} D^{\mu\alpha}g_{\alpha\beta}D^{\beta\gamma}\right) g_{\gamma\sigma}\delta\theta^{\sigma}(\lambda=0)+O(\lambda^{-3})\\
   &=-\frac{1}{\lambda}\left(D^{\mu\alpha}g_{\alpha\beta}\delta^\beta_{\sigma}-\frac{1}{\lambda} D^{\mu\alpha}g_{\alpha\beta}D^{\beta\gamma}g_{\gamma\sigma}\right) \delta\theta^{\sigma}(\lambda=0)+O(\lambda^{-3})\\
   &=-\frac{1}{\lambda}D^{\mu\alpha}g_{\alpha\beta}\left(\delta^\beta_{\sigma}-\frac{1}{\lambda} D^{\beta\gamma}g_{\gamma\sigma}\right) \delta\theta^{\sigma}(\lambda=0)+O(\lambda^{-3})\\
   &=-\frac{1}{\lambda}D^{\mu\alpha}g_{\alpha\beta}\delta\theta^{\beta}(\lambda=0)+O(\lambda^{-2}).
 \end{align}

 We see that for large $\lambda$ the amplitude of the step $\delta\theta$ becomes progressively smaller as $\lambda^{-1}$.
  \end{widetext}
\acknowledgments
This work was performed under the auspice of the U.S. Department of Energy by Los Alamos National Laboratory under 89233218CNA000001. Research reported in this publication was supported by the U.S. Department of Energy LDRD program at Los Alamos National Laboratory (Project No. 20240004DR).
\bibliography{references.bib}

\end{document}